\documentclass[preprint,aps,prl,amssymb,nobibnotes,letterpaper,superscriptaddress]{revtex4-1}

\usepackage{graphicx}
\usepackage[utf8]{inputenc}
\usepackage{hyperref}% use for hypertext links, including those to external documents and URLs

\begin{document}
\newcommand{\PSI}{Swiss Light Source, Paul Scherrer Institute, 5232 Villigen-PSI, Switzerland }
\newcommand{\ETHTh}{Materials Theory, ETH Z\"urich, 8093 Z\"urich, Switzerland}
\newcommand{\MPI}{Max Planck Institute for the Structure and Dynamics of Matter, CFEL, 22761 Hamburg, Germany } \newcommand{\ETHq}{Institute for Quantum Electronics, ETH Z\"urich, 8093 Z\"urich, Switzerland }
\newcommand{\FHIpc}{Department of Physical Chemistry, Fritz Haber Institute of the Max Planck Society, 14195 Berlin, Germany}
\newcommand{\FHIic}{Department of Inorganic Chemistry, Fritz Haber Institute of the Max Planck Society, 14195 Berlin, Germany}
\newcommand{\Bern}{Department of Chemistry and Biochemistry, University of Bern, 3012 Bern, Switzerland}

%\preprint{}

\setlength{\skip\footins}{4.5mm}
\renewcommand{\footnoterule}{}
\title{Ultrafast relaxation dynamics of the antiferrodistortive phase in Ca doped SrTiO$_3$}

\author{M. \surname{Porer}}
\email[]{Michael@Porer.org}
\affiliation{\PSI}

%\author{Michael \surname{Fechner}}
\author{M. \surname{Fechner}}
\affiliation{\MPI}
\affiliation{\ETHTh}

%\author{Elisabeth \surname{Bothschafter}}
\author{E. \surname{Bothschafter}}
\affiliation{\PSI}

%\author{Laurenz \surname{Rettig}}
\author{L. \surname{Rettig}}
\affiliation{\PSI}
\affiliation{\FHIpc}

%\author{Matteo \surname{Savoini}}
\author{M. \surname{Savoini}}
\affiliation{\PSI}
\affiliation{\ETHq}

%\author{Vincent \surname{Esposito}}
\author{V. \surname{Esposito}}
\affiliation{\PSI}

%\author{Jochen \surname{Rittmann}}
\author{J. \surname{Rittmann}}
\affiliation{\PSI}

\author{\mbox{M. \surname{Kubli}}}
\author{M. J. \surname{Neugebauer}}
\author{E. \surname{Abreu}}
\author{T. \surname{Kubacka}}
\author{T. \surname{Huber}}
\author{G. \surname{Lantz}}

\affiliation{\ETHq}

\author{S. \surname{Parchenko}}
\author{S. \surname{Gr\"ubel}}
\affiliation{\PSI}

\author{A. \surname{Paarmann}}
\affiliation{\FHIpc}
\author{J. \surname{Noack}}
\affiliation{\FHIic}

\author{P. \surname{Beaud}}
\author{G. \surname{Ingold}}
\affiliation{\PSI}

\author{U. \surname{Aschauer}}
\affiliation{\Bern}

\author{S. L. \surname{Johnson}}
\affiliation{\ETHq}
\author{U. \surname{Staub}}
\email[]{Urs.Staub@psi.ch}
\affiliation{\PSI}

\date{\today}

\begin{abstract}
The ultrafast dynamics of the octahedral rotation in Ca:SrTiO$_3$ is studied by time resolved x-ray diffraction after photo excitation over the band gap. By monitoring the diffraction intensity of a superlattice reflection that is directly related to the structural order parameter of the soft-mode driven antiferrodistortive phase in Ca:SrTiO$_3$, we observe a  ultrafast relaxation on a 0.2 ps timescale of the rotation of the oxygen octahedron, which is found to be independent  of the initial temperaure despite large changes in the corresponding soft-mode frequency. A further, much smaller reduction on a slower picosecond timescale is attributed to thermal effects. Time-dependent density-functional-theory calculations show that the fast response can be ascribed to an ultrafast displacive modification of the soft-mode potential towards the normal state, induced by holes created in the oxygen 2p states.	
\end{abstract}

%\keywords{}

\maketitle
Understanding the dynamics and speed limits of structural and/or electronic symmetry breakings is fundamental for possible applications in ultrafast data storage. Most studies in condensed matter systems are concerned with electronically driven phase transitions including either charge \cite{Rohwer2011, Beaud2014} and orbital orders, \cite{Beaud2014,Tomimoto2003} charge density wave order \cite{Moehr-Vorobeva2011,Porer2014}, magnetic phase transitions \cite{Kirilyuk2010,Koopmans2010,Ehrke2011,Johnson2012} or magnetization reversal \cite{Ostler2012,LeGuyader2015}. It is crucial to determine the timescale of these processes as well as the speed at which the crystal structure follows ultrafast modifications of electronic (magnetic) order in the time domain. This has been achieved for correlated materials such as e.g. manganites \cite{Beaud2014} and VO$_2$ \cite{Cavalleri2001,Morrison2014} by using ultrafast x-ray and electron diffraction. 

\par
Purely structural phase transitions, however, are solely governed by interactions within the phonon system \cite{Cowley1980} without any electronic or spin modulation as a driving force. For such transitions, very little is known about the dynamics induced by an electronic excitation.

In the common picture of the two-temperature model \cite{Shah1999}, phonon driven distortions might be expected to relax on a timescale of a few ps determined by the rate of heat transfer from the electronic to the lattice system. This can be compared to phase transitions driven by coupling to an electronic order parameter, where the structural relaxation is on the order of a few hundred femtoseoconds or faster \cite{Beaud2014}.
   
\par
Pristine SrTiO$_3$ is an insulating simple cubic perovskite at room temperature and undergoes a prototypical, purely structural phase transition from cubic (spacegroup Pm-3m) to tetragonal symmetry (spacegroup I4/mcm) below $T_c = 105\, \mathrm{K}$. This transition is solely driven by softening of a mode at the zone boundary \cite{Shirane1969,Riste1971}. A polar phonon mode softens as well, however, the system remains paraelectric down to the lowest temperatures \cite{Kiat1996} because quantum fluctuations \cite{Mueller1979} prevent a ferroelectric phase transition. 
\par
The antiferrodistortive structural modulation manifests itself by a rotation of the TiO$_6$ octahedra by an angle $\varphi$ (inset of Fig. 1) \cite{Shirane1969,Riste1971} at a wavevector of q=R (we keep the Pm-3m symmetry for the subsequent discussion). As  $\varphi$ is small, in equilibrium it can be seen as the order parameter of the transition [Mueller]. The modulation gives rise to x-ray superlattice (SL) reflections with intensity approximately proportional to the square of the atomic displacement $u$ of the I4/mcm oxygen 8h sites and to the square of  $\varphi$ (see supplementary material).
\par
Here we employ ultrafast X-ray diffraction to monitor the structural order parameter of SrTiO$_3$ following ultrafast optical injection of e-h pairs. We find a fast partial relaxation on a timescale of 0.2 ps. We identify these dynamics as being non-thermally driven by changes of the phonon-potentials induced by oxygen 2p hole doping. The increase of average lattice temperature is found to play a minor role. 
\par
The static properties of commercially available single crystals of $\mathrm{Sr}_{0.97}\mathrm{Ca}_{0.03}\mathrm{TiO}_3$ \cite{SurfaceNet} were characterized by hard x-ray diffraction at the Materials Sciences beamline of the Swiss Light Source (SLS) at the Paul Scherrer Institute \cite{Willmott2013}, and by Raman spectroscopy of the octahedral soft mode. We obtain a critical temperature of $T_c = 280\, \mathrm{K}$ and a soft-mode frequency of \mbox{2.5 THz} at \mbox{$T$ = 100 K} (Fig. S1). Isovalent substitution of Sr with Ca in SrTiO$_3$ leads to an increase of $T_c$, but does not change the symmetry of the distorted phase for doping levels below 6\% \cite{Mishra2005}. Our choice of 3\% of Ca-doping enables us to access the distorted phase via cooling with liquid N$_2$.

\par

\begin{figure}[t]
\includegraphics[width=0.5\columnwidth]{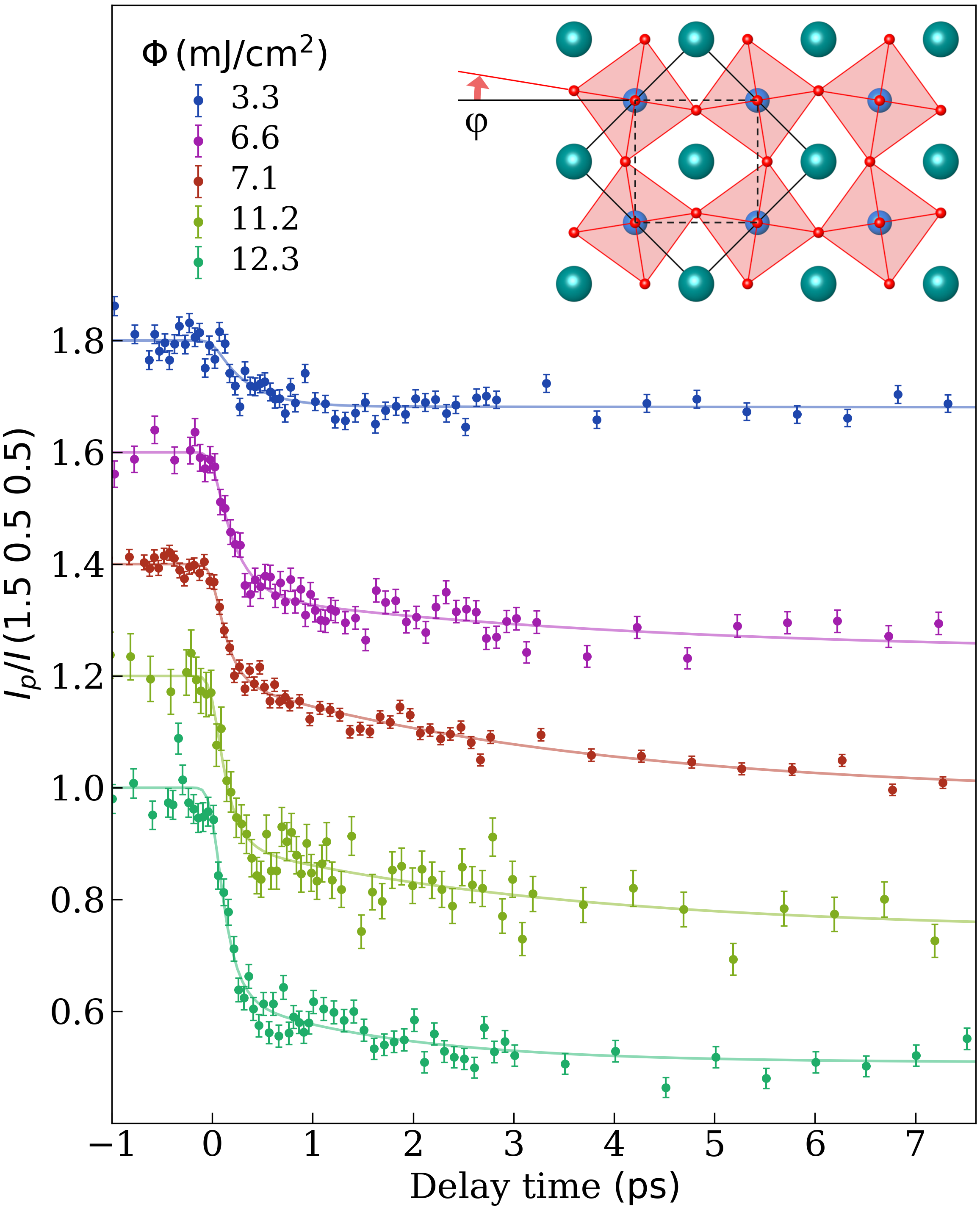}
 \caption{Transient relative x-ray intensity of the (1.5 0.5 0.5) superlattice reflection of $\mathrm{Sr}_{0.97}\mathrm{Ca}_{0.03}\mathrm{TiO}_3$ upon above bandgap excitation with 40 fs pulses centered around 4.66 eV at a temperature of 100 K. For clarity, a constant vertical offset of 0.2 separates traces recorded for different excitation fluences $\phi$. Error bars indicate x-ray photon counting statistics. Inset: SrTiO$_3$ crystal structure as seen along the c-axis (Visualized using VESTA \cite{Momma2011}). $\varphi$ measures the antiferrodistortive rotation of the oxygen octahedra (exaggerated) and represents the order parameter. The dashed/solid black line frames a cubic/tetragonal unit cell of the high/low temperature phase along the respective in-plane axes.
 	\label{fig1}}
\end{figure}
\begin{figure}[t]
	\includegraphics[width=0.5\columnwidth]{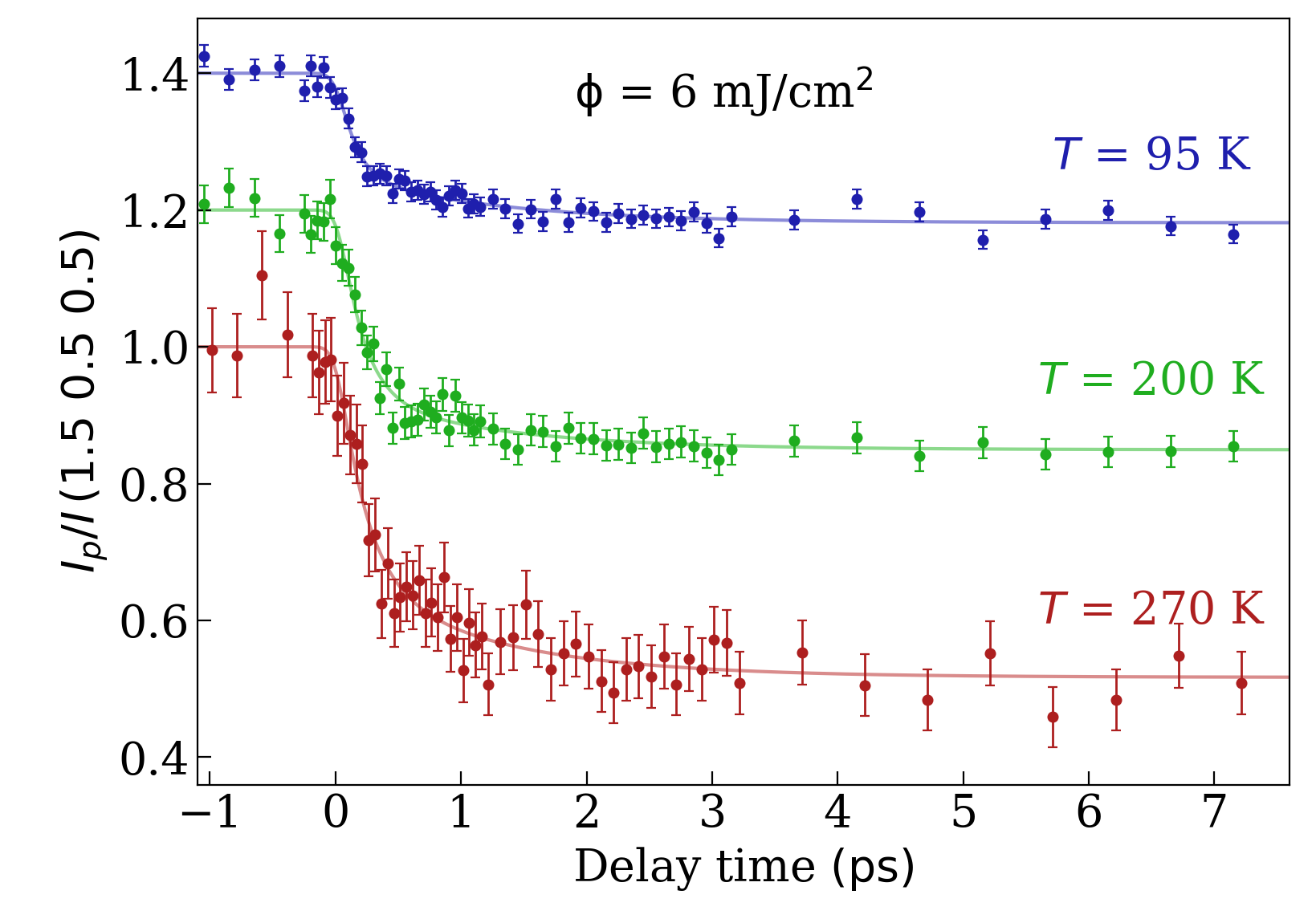}
	\caption{Dynamics of the (1.5 0.5 0.5) superlattice reflection measured for temperatures approaching $T_c = 280\,\mathrm{K}$  at an excitation fluence of \mbox{$\phi$ = 6 mJ/cm$^2$}.  
		\label{fig2}}
\end{figure}

Approximately 120 fs long x-ray pulses with 7.1 keV energy and 2 kHz repetition rate, available at the FEMTO slicing facility at SLS \cite{Beaud2007}, were used to probe the crystal structure of $\mathrm{Sr}_{0.97}\mathrm{Ca}_{0.03}\mathrm{TiO}_3$ after photoexcitation. A grazing incidence angle of 0.37$\pm$0.05$^\circ$ was chosen to limit the x-ray probe depth (intensity) to $\leq$ 60 nm.  The ~110 fs p-polarized excitation pulse is centered around an energy of 4.66 eV to overcome the direct band gap (3.75 eV \cite{Benthem2001}) leading to an excitation depth of 18 nm. This results in an average photodoping of the probed volume of ~0.01 eh-pairs per cubic unit cell for 1 mJ/cm$^2$ incident fluence. The crystal was cooled with a N$_2$ cryo-blower.

\par
Figure 1 shows the normalized intensity of the \mbox{(1.5 0.5 0.5)} superlattice reflection as a function of the pump-probe delay for a series of excitation fluences $\phi$ taken at \mbox{$T$ = 100 K}. A sudden reduction immediately after excitation is visible, followed by a subsequent slower decay during a few ps.  For quantitative extraction of the time constants of the order parameter dynamics, we fit the normalized intensity by two exponential decays,

\begin{displaymath}
I=\left( 1-\Theta(t\!-\!t_0) \left[A_1\left(1\!-\!\mathrm{e}^{-\frac{t\!-\!t_0}{\tau_1}}\right) +A_2\left(\!1-\!\mathrm{e}^{-\frac{t\!-\!t_0}{\tau_2}}\right)\right]\right)^2
\end{displaymath}

convolved with a 120 fs full-width at half-maximum Gaussian to account for overall experimental time-resolution.
$\Theta$ is the Heaviside function, and $A_1/\tau_1$ and $A_2/\tau_2$ are the amplitude/decay constant values of the fast and slow decay components, respectively. Due to the mismatch between UV excitation and x-ray probe depth mentioned above, we base our discussion and conclusions solely on the decay constants. The fits result in a fast and slow decay constant of $\varphi$ of approximately \mbox{0.2 ps} and \mbox{$>1$ ps}, respectively (see supplementary material). 
\par
The slower decay component corresponds to the expected time scale for transfer of heat from the electronic system \cite{Shah1999} to the lattice, so this component of the relaxation seems to be likely due to a simple temperature increase of the phonon system with the distortion following adiabatically. It is also possible that it is influenced by energy transport between the more exited near-surface region and the deeper regions of the sample.
\par
The faster time scale is on the order of a half cycle period of the rotational mode at \mbox{100 K} (2.5 THz, see supplementary material). However, when approaching $T_c$ at \mbox{270 K} the fast dynamics does not change significantly, which is in contrast to the increasing equilibrium soft mode period with temperature (\mbox{0.8 ps} at \mbox{$T=270\,\mathrm{K}$}, see Fig. 2 and Fig. S2(b)). This points towards a non-adiabatic structural dynamics that is being driven by a faster than thermal modification of the structural soft-mode potential.
\par
In principle, changes of the Debye-Waller factor imposed by optically induced lattice disordering can account for a strong reduction in the intensity of a Bragg peak on comparably fast timescales \cite{Hillyard2007,Lindenberg2005}. In our case, a drop by 40\% of the probed reflection would imply an average thermal isotropic atomic displacement by more than 60\% of the cubic lattice constant. According to the Lindemann criterion, this would imply melting of the crystal lattice, which excludes such a scenario. As discussed in the supplementary material, we can also exclude a rapid order-disorder transition of the structural superlattice itself, i.e. a selective thermal dislocation of the 8h oxygen sites of the I4/mcm cell along the soft-mode coordinate.

\par
For metallic systems with strong electron-lattice interaction, it has been shown that an ultrafast modification of the electronic population distribution can change the phonon potentials and non-thermally drive structural dynamics along zone-centered $A_{1g}$ optical modes \cite{Fritz2007}. To explore whether photodoping across the band gap of an insulator can induce a non-thermal relaxation dynamics of a soft-mode driven distortion via a comparable mechanism, we model the photodoping process in SrTiO$_3$ from first principle calculations. \par
We first look into the effect of static doping on the phonon spectrum using density functional theory (DFT) \cite{Kresse1996} (VASP code). As a 3\% Ca substitution does not qualitatively change the properties of the octahedral phase transition, we calculated the effect of hole doping on the phonon dispersion for pure SrTiO$_3$ following the approach for electron doping of Ref. \cite{Cancellieri2016}

\begin{figure}[t]
	\includegraphics[width=0.5\columnwidth]{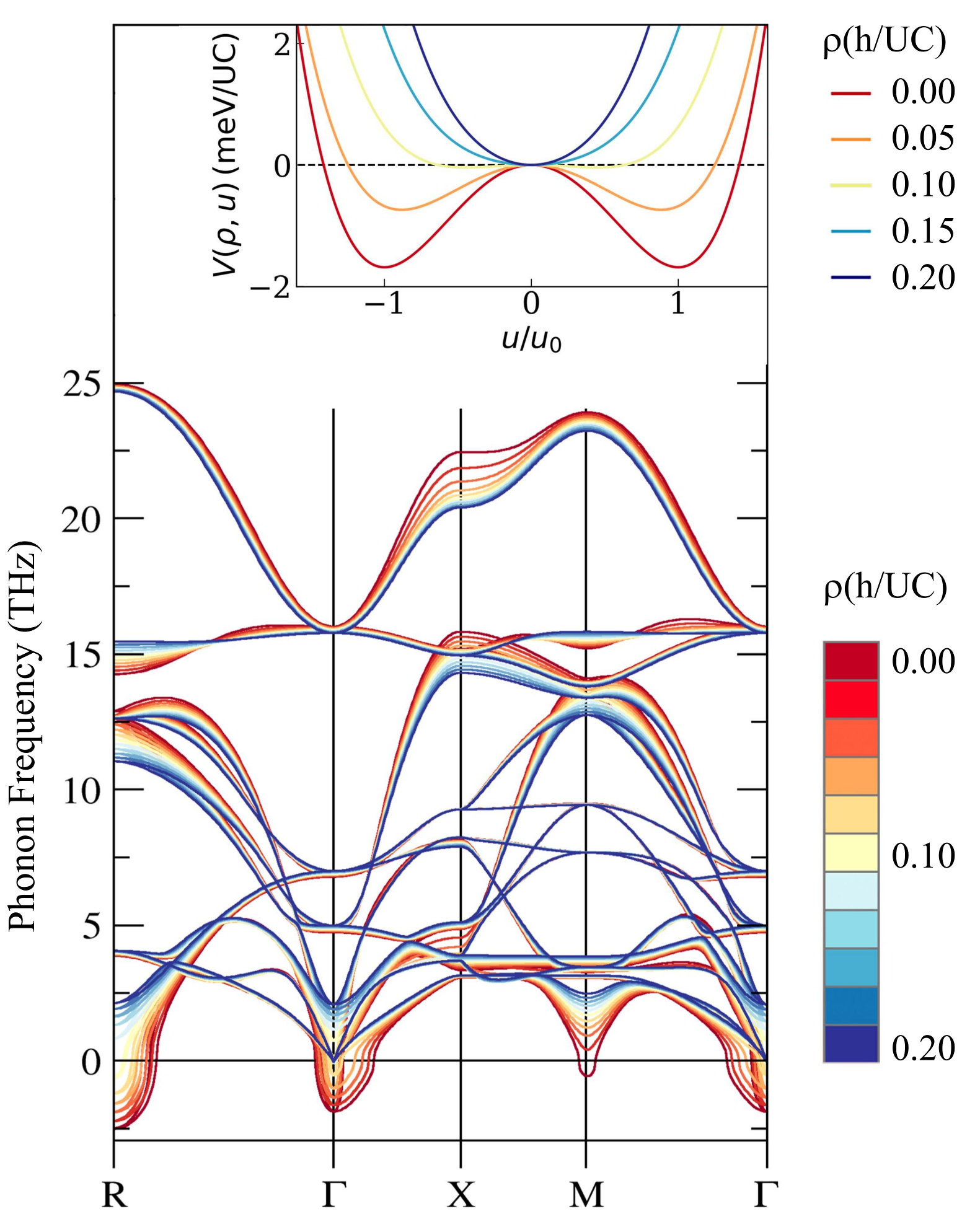}
	\caption{DFT Phonon dispersion of cubic SrTiO$_3$ for a series of hole dopings (not including LO-TO splitting). Inset: Calculated energy change per SrTiO$_3$ cubic unit cell as a function of oxygen displacement $u/u_0$ along the in-plane cubic crystal axes resulting from the octahedral rotation at select hole dopings.
		\label{fig3}}
\end{figure}

\par
In Fig. 3 we show the phonon dispersion of SrTiO$_3$ for different hole doping concentrations. Within the harmonic approximation, phonon modes with an imaginary frequency represent structural instabilities of a system. Without hole doping the spectrum exhibits phonon bands with imaginary frequencies (shown as 'negative' frequencies) around the R and $\Gamma$ points. The instability at R represents the rotational low temperature ground state of SrTiO$_3$, whereas the one at $\Gamma$ indicates its polar instability. Introduction of holes significantly shifts the lowest phonon bands upwards in energy, such that imaginary frequencies vanish at a critical concentration of 0.1 holes per unit cell. Consequently, hole doping suppresses rotational and polar instabilities and instead favors a cubic undistorted structure, in agreement with literature \cite{Uchida2003}. Electron doping, in contrast, does not induce this effect as rotational modes are almost unaffected \cite{Cancellieri2016}.
\par
Next, we quantify the potential energy landscape resulting from a modulation of the phonon at the R point by a harmonic double-well potential of the form
\begin{equation}
V(u,\rho)=\frac{\Omega(\rho)^2}{2}u^2+\frac{\kappa(\rho)^2}{4}u^4
\end{equation}
where $\Omega(\rho)$ is the doping dependent phonon frequency and the second term with $\kappa$ represents a higher order repulsive force. Minimizing Eqn. (1) gives the structural ground state, i.e. the octahedral rotational angle for each doping. The inset of Fig. 3 shows $V(\rho)$ for a series of doping values as a function of the oxygen displacement u relative to the optimal displacement at zero doping $u_0$. The transformation of the double well to a single well potential for $\rho$ above 0.1 h/UC reflects the stabilization of the cubic structure by doping.

\par

To link the structure to the dynamic photodoping process, we replace the static doping $\rho$ by a time-dependent hole concentration $\rho(t)$ which becomes a parametric driving force in Eqn. (1) that displaces $u$. The structural dynamics is then given by the equation of motion

\begin{equation}
\ddot{u}(t)+2\gamma\dot{u}(t)+\nabla_u V(u(t),\rho(t))=0
\end{equation}

in which we account by $\gamma\dot{u}(t)$ for finite phonon lifetimes. We note that within this approach the interplay between $\rho(t)$ and $V(u,\rho)$ ultimately determines the dynamics of the structural distortion.

\par

To account for a realistic transient doping $\rho(t)$ we next perform computations utilizing time-dependent DFT \cite{elk}. We hereby describe the photodoping process by the light pulse by including a time-dependent vector potential within our calculation. $\rho(t)$  is then determined by integrating the time-dependent density of holes at the top of the valence band, between $E_F-0.2\mathrm{eV}$ and $E_F$ (which is in the vicinity of R), to capture only the hole states considered in the static calculation. Details of our calculations are presented in the supplementary part. Figure 4 shows an example timetrace of $\rho(t)$.

\begin{figure}[t]
	\includegraphics[width=0.5\columnwidth]{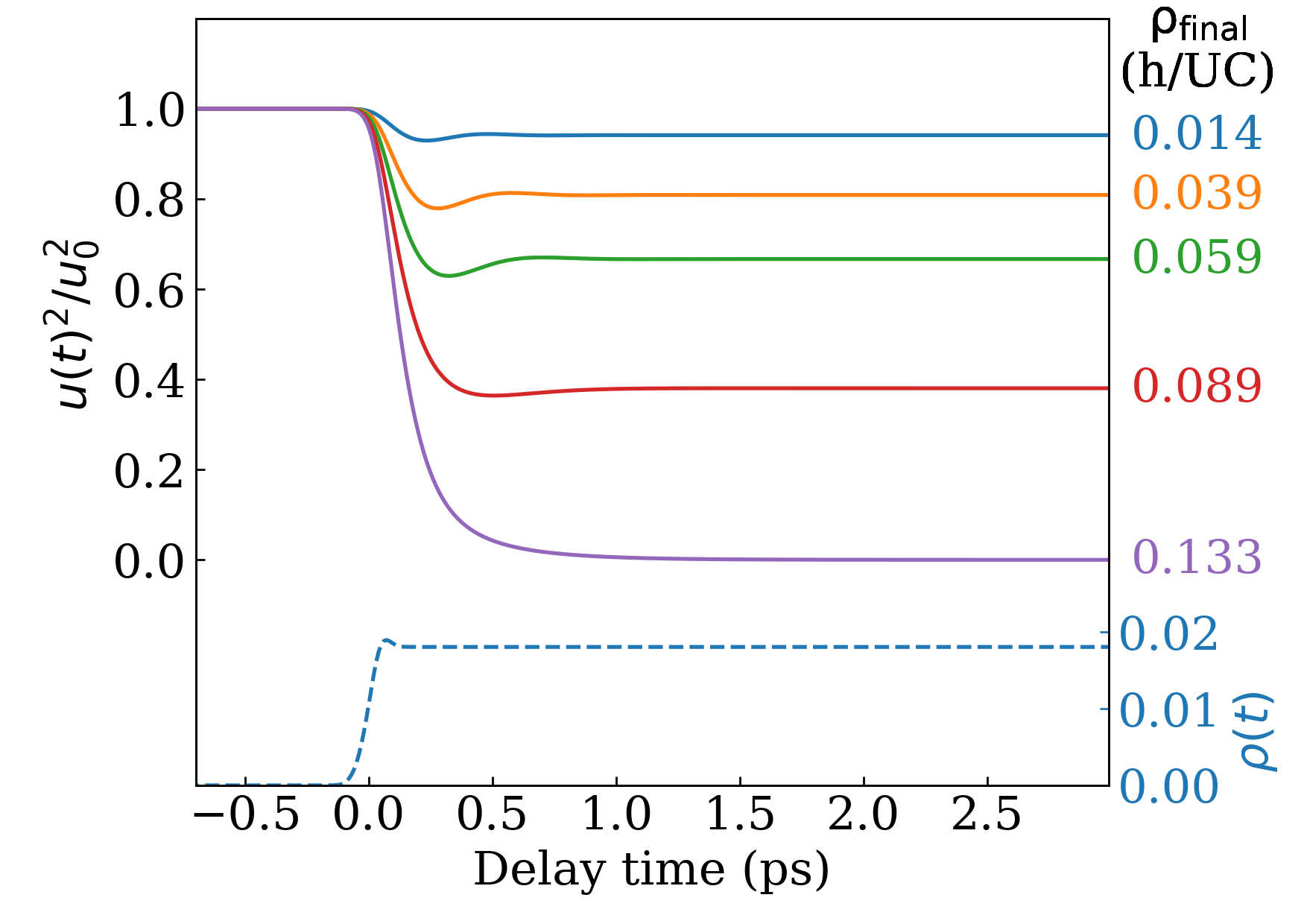}
	\caption{Right y-axis: TDDFT simulation of the transient hole population $\rho(t)$ for excitation with a ~110 fs field pulse centered 200 meV above the DFT bandgap (see Fig. S5). Left y-axis: Relative scattering intensity of superlattice peaks associated with the octahedral distortion, calculated via Eq. 2 for a parametric change of the double well potential of Fig. 3 by $\rho(t)$. The curves are labelled by the doping level $\rho_{final}$ reached after the excitation process.\label{fig4}}
\end{figure}
\par
Having determined the transient doping $\rho(t)$ allows us to calculate the dynamics of the mode coordinate $u(t)$ by solving Eqn. (2). We show $u^2(t)$ as a function of time in Fig. 4. Here we use the damping $\gamma$ of the soft mode obtained by ref. \cite{Kohmoto2006} extrapolated to \mbox{$T$ = 120 K}. Note that the experiment captures a superposition of excitation intensities due to the mismatch between excitation and probe depths (supplementary material). For a fluence of 3.3 mJ/cm$^2$, the calculated doping level close to the surface is 0.11 h/UC which is thus already sufficient to stabilize the cubic phase in part of the probed volume. In figure S6 we show the calculated transient reflection intensities with the influence of the profile mismatch taken into account using the model described in ref. \cite{Huber2014}. The amplitude of the intensity drop obtained by the model is in a good overall agreement with the experimental amplitude of the fast decay component.
\par
Most notably, we find that the timescale of the initial fast relaxation is qualitatively reproduced (Fig. 4 and Fig. S6). This shows that ultrafast modification of the soft-mode potential via photodoping of oxygen 2p valence holes can explain the observed fast relaxation of the structural order parameter. The following slower decay in the experimental data is not reproduced due to the absence of electron-phonon scattering in the model. Due to the nanosecond lifetime of photodoped e-h pairs \cite{Yasuda2008}, recombination processes are not expected to be relevant on the timescale of interest.     
\par
In conclusion, our time resolved study of the order parameter of a soft-phonon induced purely structural symmetry breaking transition demonstrates that optical excitation of the electronic system can induce a relaxation of the structural order parameter faster than expected for lattice heating. Our results show that photodoping can modify the structural soft-mode potential and drive the order parameter, despite the equilibrium phase transition is not based on an electronic mechanism.

\end{document}